\documentstyle[12pt,aasms]{article}

\def\plotfiddle#1#2#3#4#5#6#7{\centering \leavevmode
    \vbox to#2{\rule{0pt}{#2}}
    \includegraphics{#1}}

\def\etal{{\it et~al.\ }}
\tightenlines

\begin{document}

\title{The Luminosity Function for $L>L^*$ Galaxies at $z > 3$ }

\author{Matthew A. Bershady\altaffilmark{1,6} and Steven R.
  Majewski\altaffilmark{2,6}, \\ David C. Koo\altaffilmark{3,6}
  Richard G. Kron\altaffilmark{4,6}, and Jeffrey A.
  Munn\altaffilmark{5,6}}

\altaffiltext{1}{Dept. of Astronomy \& Astrophysics, Pennsylvania
  State University, University Park, PA 16802, and Dept. of Astronomy,
  University of Wisconsin, Madison, WI 53706 (mab@astro.wisc.edu)}

\altaffiltext{2}{Dept. of Astronomy, University of Virginia,
  Charlottesville, VA, 22903-0818
  (srm4n@didjeridu.astro.virginia.edu)}

\altaffiltext{3}{University of California Observatories/Lick
  Observatory, Board of Studies in Astronomy \& Astrophysics,
  University of California, Santa Cruz, CA 95064 (koo@ucolick.org)}

\altaffiltext{4}{Fermi National Accelerator Laboratory, MS 127, Box
  500, Batavia, IL 60510 (rich@oddjob.uchicago.edu)}

\altaffiltext{5}{U.S. Naval Observatory, Flagstaff Station, P.O. Box
  1149, Flagstaff, AZ 86002-1149 (jam@nofs.navy.mil)}

\altaffiltext{6}{Visiting Astronomer, Kitt Peak National Observatory,
  National Optical Astronomy Observatories, which is operated by the
  Association of Universities for Research in Astronomy, Incorporated,
  under cooperative agreement with the National Science Foundation}

\vskip 2in
\centerline{Accepted for publication in The Astrophysical Journal Letters}

\newpage
\begin{center}
{\large\bf Abstract}
\end{center}

Through use of multiband ($U, B_J, R_F, I_N$) photometry
we have isolated high redshift (3.0$<$z$<$3.5) galaxy candidates in a
survey of 1.27 deg$^2$ to $R_F = 21.25$ and a survey of 0.02 deg$^2$
to $R_F = 23.5$. Our pool of candidates constrains the nature of the
$3.0 < z < 3.5$ luminosity function over the range $L^* \lesssim L
\lesssim 100 L^*$, if we grant a similar level of completeness to
these data as for very faint samples (to $R = 25.5$) selected in a
similar fashion. Our constraints agree with the high redshift sky
density at $R_F = 20.5$ estimated from Yee \etal's (1996)
serendipitous discovery of a bright, $z=2.7$ galaxy, as well as the
density at $R_F \approx 23$ by Steidel \etal (1996b). We strongly rule
out -- by more than two orders of magnitude at $M_{R_F} = -25$ -- the
$L > L^*$ luminosity function for $z = 3-5$ galaxies obtained by a
photometric redshift analysis of the Hubble Deep Field (HDF) by Gwyn
\& Hartwick (1996). Our results at $R_F \approx 23$ are more
consistent with the photometric redshift analysis of the faint HDF
galaxies by Sawicki \& Yee (1996), but our present upper limits at the
brightest magnitudes ($R_F<21.5$, $M_{R_F} < -24$) allow more generous
volume densities of these super-$L^*$ galaxies.


\dates

\clearpage

\section{Introduction}

When deep imaging surveys revealed a significant population of blue
objects (well in excess of no-evolution models) at $B > 23$, it was
initially thought (Tyson 1988, Cowie 1988, Cowie {\it et al.} 1988)
that this could be the signature of ``primeval galaxies'' (PGs) --
counterparts to present day $\approx L^*$ galaxies undergoing an
initial, extremely bright burst of star formation at high $z$
(Partridge \& Peebles 1968). Ever deeper redshift surveys (now $B \sim
24$, Songaila {\it et al.} 1994, Glazebrook {\it et al.} 1995),
however, revealed galaxies only to $z < 0.8$, and it became evident
that galaxies with redshifts as high as 3 were not likely to represent
a substantial fraction of the galaxies, even to $B \sim 25$ (Koo {\it
  et al.} 1996).

The conspicuous paucity of faint, high $z$ galaxies had already
been shown by two studies of faint galaxy {\it colors}.
At $2.7<z<3.4$, galaxies should exhibit a particularly red $U-B$, 
compared to a rather blue color at longer wavelengths, because
of the presence of the Lyman limit in the $U$ passband. The first
application of this test by Koo, Kron \& Majewski (see Majewski 1988,
1989) demonstrated the number of $B < 24.5$ galaxies showing
the expected $z>3$ color signature to be negligible -- $<$ 1\%.  With
deeper data, Guhathakurta, Tyson \& Majewski (1990) showed
that the number of galaxies to $B \sim 27$ showing the expected $z>3$ 
color signature was no more than 7\%, and likely $<1$\% of galaxies.
 
More recently, Steidel \etal (1995 and references therein, ``S95'')
have repeated the ``Lyman limit imaging'' experiment over $\sim0.03$
deg$^2$ to their $\Re = 25.5$. They confirm the relatively low
surface-density of high $z$ candidates, and with the Keck 10-m
have spectroscopically verified 22 of 37 color-selected candidates are
indeed at $3 < z < 3.5$ (Steidel \etal 1996b, ``S96''). All galaxies
in S96 have $\Re = $23.7-25.5, implying luminosities near present day
$L^*$ and slightly brighter. From their data, S96 estimate the
comoving space density of these objects to be approximately 1/2 that
of present day $L>L^*$ galaxies. Both Steidel \etal (1996a) and
Lowenthal \etal (1997) find comparable results in the same magnitude
and redshift range in the Hubble Deep Field (HDF), despite more
liberal color selection in the latter survey.

The numbers of objects much brighter than $L^*$ is less well
constrained.  Yee \etal (1996) have discovered serendipitously 
a ``normal'' (i.e., neither AGN nor radio), $V=20.5$ galaxy at
$z=2.7$. Though super-luminous at $M_{R} - 5 \ {\rm log} \ h \approx -25$
(q$_0$=0.5, $h$ = H$_0$/100), this galaxy is spectroscopically similar
to the S96 galaxies. Based on one galaxy in their 0.66 deg$^2$ survey,
Yee \etal estimate the density of such objects at $R\sim20$ is
$10^{0\pm1}$ deg$^{-2}$.

Meanwhile, Gwyn \& Hartwick (1996, ``GH'') attempted to determine {\it
photometric} redshifts for galaxies in the HDF and
claim dramatic changes in the galactic luminosity function
($\Phi(M)$) from $0<z<5$ with $\Phi(M)$ becoming flat between $-24\leq
M_B \le -15$ for $3<z<5$. They predict a substantial {\it abundance}
of galaxies up to $M^*-4$ at high $z$. In stark contrast, the
photometric redshift analysis of the HDF by Sawicki \etal (1996,
``SLY'') finds a more prosaic $z=3-4$ luminosity function, adequately
described by a Schechter function with $\alpha = -1.3$ and ${\phi}^* =
0.023 h^3$ Mpc$^{-3}$. An extrapolation of this
function predicts many orders of magnitude less high $z$ galaxies
at $R_F=21$ than GH. Yet another photometric analysis of the HDF by
Mobasher \etal (1996) suggests strong luminosity function evolution to
$z = 3$, and implies numbers of bright $z=3$ galaxies intermediate
between GH and SLY. While all of these groups suggest that in the HDF
they are seeing the formation of $L \ge L^*$ galaxies at high $z$,
there seem to be vast differences in the implied nature of the
luminosity function, especially for bright galaxies.

The range in HDF results may be attributable to the substantial
uncertainty in the application of photometric redshifts at very faint
magnitudes. Unlike the photometric redshift study of brighter galaxies
by Connolly \etal (1996), at HDF depths there is no adequate
spectroscopic training set available for calibration. While S96 have a
handful of spectra of $z>3$ galaxies, the vast majority of objects to
$\Re\approx 25$ and beyond are without spectroscopic redshifts. Most
troubling to the interpretation and application of spectro-photometric
galaxy models is the near-degeneracy in color between particular
redshifts (e.g., at $0<z<1$ for the bluest galaxies and $z\approx 2.5$
for all galaxies; see figure 1); this plausibly produced the apparent
strong bimodality in the redshift distribution inferred by GH. Unlike
in Connolly \etal, GH and SLY do not use apparent brightness to break
such degeneracies in the redshift estimates. Moreover, as pointed out
by SLY, differences in the model spectral energy distributions (SEDs)
-- particularly in the still poorly understood rest-frame ultraviolet
where internal reddening and intervening absorption are important --
lead to substantially different results. It is important, therefore,
to check the HDF results, especially at magnitudes where spectroscopic
confirmation is feasible.

Apart from the S96 data near $L^*$ and the loose Yee \etal (1996)
constraint, there is scant spectroscopic redshift data to explore the
nature of the luminosity function of $z > 3$ galaxies.  Ironically, it
is $\Phi(M)$ at magnitudes brighter than those explored by S96 that is
most poorly defined due to a lack of reliable data; if the high redshift
$\Phi(M)$ is of prosaic form, with a steep decline toward the bright
end, much larger survey areas are required to explore the $L >> L^*$
domain than have been achieved with CCD surveys to date. We have
undertaken a large area, photometric search for bright, high $z$
galaxies.  Even without spectroscopy, we constrain the high $z$
luminosity function based on the magnitude distribution of our high
$z$ candidates. To do so, we rely on the good correspondence between
high $z$ galaxy candidates identified by S95 through similar selection
criteria and {\it bona fide} high $z$ galaxies among these candidates
as confirmed by S96.

\section{High Redshift Galaxy Search}

Our search for bright, $z>3.0$ galaxies utilizes two sets of
multicolor galaxy catalogues. The first data set consists of the
photographic catalogues generated for the Kitt Peak Galaxy Redshift
Survey (KPGRS; Munn {\it et al.} 1997) and faint quasar surveys (Kron
{\it et al.} 1992). These catalogues cover four separate regions of
the sky totaling 1.27 deg$^2$ with photometry from sky-limited, Mayall
4-m photographic plates in the $UB_JR_FI_N$ passbands\footnote{Our
  $U$ band is virtually identical to the standard photoelectric $U$
  (Koo 1985).  Steidel \& Hamilton's (1993) $U_n$ is about 100 \AA \ 
  bluer, and $U_{F300W}$ for the HDF is about 700 \AA \ bluer.}
calibrated with deep CCD photometric sequences (e.g. Majewski {\it
  et al.} 1994). While these catalogues reach to $R_F=23$, we choose
here a conservative catalogue limit of $R_F=21.25$, where random
errors in $B_J, R_F$ and $I_N$ are at most 0.3 mag (smaller than the
color difference between our $z > 3$ selection thresholds and the
locus of low $z$ galaxies).

The panels in Figure 1 show the progression of $U-B_J$,
$B_J-R_F$ and $R_F-I_N$ galaxy colors with redshift.
The iso-$z$ loci for different galaxy types were generated
with Bruzual \& Charlot's
(1993) models for a range of star-formation histories\footnote{ Colors
  are for evolving and non-evolving model galaxies with 16 Gyr ages at
  $z=0$ (q$_0$=0.1, H$_0$=50 km s$^{-1}$ Mpc$^{-1}$, $\Lambda$=0),
  Salpeter initial mass, and $0.01<\mu<0.95$, where $\mu$ is the
  fraction of galactic mass in stars after 1 Gyr of star formation.},
plus observed elliptical and starburst (using the galaxy N4449)
spectral energy distributions (SEDs). Model and observed SEDs were
reddened as a function of band and redshift to account for
intervening absorption (only), as prescribed by Madau
(1995). This process is identical to that in S96, except that we
include a broader range of SEDs. For colors of
$z>3.0$ galaxies this has little consequence, however it is important
at lower redshifts. Based on our models, dashed lines delimit the
region of each color-color diagram inhabited by $z>3.0$ galaxies. To
justify this selection, we show (top-left panel, Figure 1)
the S96 $z>3.0$ galaxies in our $UB_JR_F$ system using
the transformations in Majewski (1992) and Steidel \& Hamilton (1993);
the symbols are coded for those objects S96 classified as ``robust''
($U$-band drop-outs falling within the selection boundary based on
1-$\sigma$ limits), and ``marginal'' (non-``robust'' objects within the
selection boundary). We also show (top-right panel,
Figure 1) all known $2.7<z<3.0$ and $z>3.0$ QSOs in our fields (Kron
\etal 1992).

The middle and right panels of Figure 1 show our $R_F<21.25$ sample.
We are interested in setting upper limits on the numbers of bright,
high $z$ galaxies; our selection algorithm reflects a liberal
acceptance threshold that sets a conservative upper limit on $\Phi(M)$
while maintaining reasonable reliability. As a first acceptance
criterion, we adopt a similar selection function to S96 -- objects
with red $U-B_J$ and blue $B_J-R_F$ colors, as illustrated in the top row of
Figure 1. However, our models show the $U-B_J, R_F-I_N$ diagram (middle row,
Figure 1) affords a much cleaner separation of high $z$ galaxies from
the low $z$ locus. We accept high $z$ galaxy candidates from this
diagram as well. Galaxies selected in the ($U-B_J$, $B_J-R_F$) and
($U-B_J, R_F-I_N$) diagrams need not be the same. For example, a high
$z$ galaxy might be missed in the $U-B_J, B_J-R_F$ plane if the line
of sight to that object passes through a sufficient number of neutral
hydrogen clouds for significant suppression of the observed $B_J$ flux
(with both $U-B_J$ and $B_J-R_F$ affected).

The $B_J-R_F, R_F-I_N$ diagrams in the bottom panels of Figure 1
reveal the high $z$ locus is not as well separated as in the other
diagrams. As a compromise between completeness and low $z$
contamination (reliability), we (1) adopt a more conservative color
cut, but (2) accept galaxies in this diagram only if they are {\it
  bona fide} $U$ band drop-outs. Objects selected in this way satisfy
the relevant $R_F-I_N$ color criterion in the ($U-B_J, R_F-I_N$)
diagram, but have $U-B_J$ upper limits insufficient to place them
confidently within the $z>3$ region. These objects are faint in $B_J$,
so their exclusion in the middle panel is likely due only to the
magnitude limit of the $U$ plates. Hence these are plausible $z>3$
candidates. In the $B_J-R_F, R_F-I_N$ diagrams we have the potential
to discover galaxies at redshifts even higher than 3.5, yet no such
``$B_J$-band drop-outs'' were found.

Each high $z$ candidate was inspected visually on a number of
photographic plates to ensure reliability. We find twelve {\it
  resolved} (unlikely to be either stars or QSOs) $z>3.0$ candidates
between $19.25<R_F<21.25$. We also find 20 {\it unresolved} sources
with $R_F<21.25$; three are spectroscopically identified as $z>2.9$
QSOs and two as stars in our QSO survey (Kron \etal 1992).  No
galaxies at $z>1$ have been identified among any of the QSO candidates
in Kron \etal (1992).

To bridge our study of the high $z$ luminosity function from $R_F =
21.25$ to the very deep S96 sample, we have generated deeper images in
two 39 arcmin$^2$ subfields of SA 57 by stacking PDS microdensitometer
plate scans of five $U$ plate images, ten $B_J$ plate images and five
$R_F$ plate images (see Majewski 1988). Candidate $z>3$ galaxies are
selected here only on the basis of $U-B_J, B_J-R_F$ as before (no
comparably deep $I_N$ image was available). A total of 11 candidates
(both stellar and nonstellar) are found (small triangles in top-middle
panel of Figure 1) to the conservative limit of $R_F = 23.5$.

\section{High Redshift Luminosity Function} 

Our $z > 3$ galaxy candidates represent the highest possible density
of bright, S96-like galaxies if we assume a similar level of
completeness in our sample as has been assumed for S96. We believe no
$z > 3$ galaxies lie within our unresolved sample to $R_F < 21.25$,
but, in the spirit of upper limits, we include discussion of this
sample here. Our survey should be {\it more} complete than, for
example, S96 since we utilize multiple combinations of colors. If
there exists a population of high $z$ galaxies not chosen by our
selection criteria, our comparison to other surveys using similar
S96-like selection criteria is still valid.

In Figure 2 we compare the various studies of the $z > 3$ luminosity
function in a cosmologically model-independent way: 

(i) The luminosity
functions of GH (in the range $3<z<5$) and SLY (in the range $3<z<4$)
are transformed into the apparent differential counts A($R_F$) for
galaxies lying in the redshift shell between 3 and 3.5, the range of
redshift to which our data apply. To do this we have adopted their
respective cosmologies to scale by the appropriate volumes and
luminosity distances, and assumed $k$-corrections for N4449. The
observed spectrum of N4449 has been extended below 1250 {\AA} using
the best-matching Bruzual \& Charlot model in the range 1250-2000 \AA.
Note that GH and SLY calculate ``the $z>3$ luminosity function''
beyond our $z=3.5$ limit, yet both studies find the space density
falling rapidly beyond $z \approx 2$. Thus our estimation of the
predicted A($R_F$) for their $\Phi(M_{R_F})$ in the {\it lowest}
redshift shell of their broader $z$ ranges provides {\it lower} limits
on the counts.  This is particularly relevant to the gross discrepancy
between our derived {\it upper limits} to A($R_F$) and the GH results
(presented as {\it lower} limits) detailed below.  

(ii) Candidate $z>3$ galaxies at faint magnitudes were compiled from
S95 and S96 in two ways. a) In S95, four of five fields had
well-defined samples and areas (i.e. excluding Q0000-263), yielding 15
candidates in 20.7 arcmin$^2$ to $\Re < 25.5$ defined as ``robust'' by
them. From S96, we counted candidate and confirmed $z>3$ galaxies in
two fields (Q0000-263 and SSA 22) to the same depth in a total area we
estimate to be 45.4 arcmin$^2$.  These tallies exclude
spectroscopically confirmed stars, QSOs, or galaxies at $z<3$. A total
of 36 robust candidates in 66.1 arcmin$^2$ are counted, 22 of which
are spectroscopically confirmed. Based on the 81\% reliability of
spectroscopically identified candidates (3 of 16 are QSOs) we scaled
the remaining unconfirmed robust candidates to derive number counts in
the $R_F$ band for $z>3$ galaxies. b) For S95, we have counted, in the
same 4 of 5 fields, additional objects not considered ``robust'' but
still within the color region believed to contain $3<z<3.5$ galaxies.
This yielded 23 candidates (including ``robust'' ones) in 20.7
arcmin$^2$.

(iii) We include the $z=2.7$ Yee \etal as a datum at $R_F\sim 20.35$
assuming one source in their survey area at this apparent brightness.

Without redshifts, we constrain the density of {\it resolved} $z>3.0$
galaxies to $\leq 10 \pm 3$ deg$^{-2}$ at $R_F<21.25$, or $\leq 22 \pm
4$ deg$^{-2}$ if we include {\it unresolved} candidates (minus the
known QSOs and stars). Our counts are in strong conflict -- at least
two orders of magnitude -- with GH at this depth. We note that the
counts of $z > 3$ galaxies in the GH analysis rivals our {\it total}
galaxy counts for $R_F<22$ (Figure 2), although if we had adopted
redder $k$-corrections, the results of the surveys would be in
somewhat better agreement.  At $22 < R_F < 23.5$, our limits are
similar to the upper limits of the more liberal set of the fainter S96
candidates ([ii]b above). The combination of the S96 upper limits and
our limits at brighter magnitudes suggests a rapid decline in the $z >
3$ luminosity function brighter than $M_R - 5 \ log \ h \approx -20.5$
to $-21$. This corresponds to the SLY \etal $M^*$ at $z=3.25$
($R_F=24.5$), or $\approx$0.75 mag brighter than the local galaxy
luminosity function. In general, our upper limits and those derived
from S95 agree with the SLY luminosity function for $23.5 < R_F <
25$. However, over the brighter range of our survey, $19 < R_F <
23.5$, our upper limits allow for a much more gradual bright-end
decline than suggested by the SLY extrapolation, and are consistent
with the Yee \etal (1996) serendipitous discovery.

While the SLY luminosity function overestimates the numbers of robust
candidates from S95 and S96, our more liberal selection from their
data is marginally consistent to $R_F \approx 25$. However, it is
critical to discriminate between the SLY suggestion of a rising faint
end of the luminosity function and the flatter faint end hinted at by
the S96 data. Upper limits on the shape of the $L < L^*$ luminosity
function could be checked via the $U$-band drop-out method in a similar
manner to what we have done here at brighter magnitudes.  While the
HDF data are appropriately deep for such an exercise, the $U_{F300W}$
of WFPC2 is much bluer than ground-based, Johnson $U$; this increases
the sensitivity of HDF data to $z \sim 2.2$ and complicates direct
comparisons. Hence there is a need for ultra-deep imaging in the
Johnson $U$ band even over relatively small fields of view.

\newpage

{\sc fig}. 1. \ -- \ $U-B_J, B_J-R_F, R_F-I_N$ \ color-color diagrams.
Left panels: the locus of non-evolving observed, model, and passively
evolving model galaxy colors (see key in middle-left panel) as a
function of redshift (labels, connected by dotted lines of constant
$z$); our $z>3$ color-selection (long-dashed lines); the $3<z<3.5$
samples from S95 and S96, transformed to our photometric system
(top-left); and the Yee \etal datum, transformed from their $g,V,r,I$
bands using relations from Fukugita \etal (1995; bottom-left).  Middle
panels: {\bf our sample} of all galaxies (points) and extended high
$z$ candidates to $R_F=21.25$ and high $z$ candidates from stacks of
photographic plate images to $R_F=23.5$. Right panels: {\bf our
  sample} of all stellar sources (points), stellar high $z$ candidates
to $R_F=21.25$, and confirmed QSOs and stars. Keys in middle and right
panels apply to all middle and right panels. Note the significant
number of our candidates in the same region of color-space as S96's
confirmed $3.0<z<3.5$ sample. We select high $z$ candidates, however,
using {\it all three} color-color diagrams.

{\sc fig}. 2. \ -- \ Differential counts of candidate and confirmed
$3<z<3.5$ galaxies from figure 1, as described in text and key.  Error
bars enclose 68.3\% confidence intervals (Gehrels 1986). Also shown
(lines) are counts for $3<z<3.5$ galaxies from luminosity functions
inferred from the HDF via photometric redshifts (GH and SLY). The
dotted portion of SLY is an extrapolation of their best-fitting
Schechter function. The absolute magnitude scale at $z=3.25$ is shown
at the top, while the change in $R_F$ with redshift for constant
luminosity and q$_0$=0.5 is shown at the bottom. {\it Total} $R_F$
galaxy counts from Kron (1980) and Majewski (1988) are shown for
reference.

\clearpage

\begin{figure}
\plotfiddle{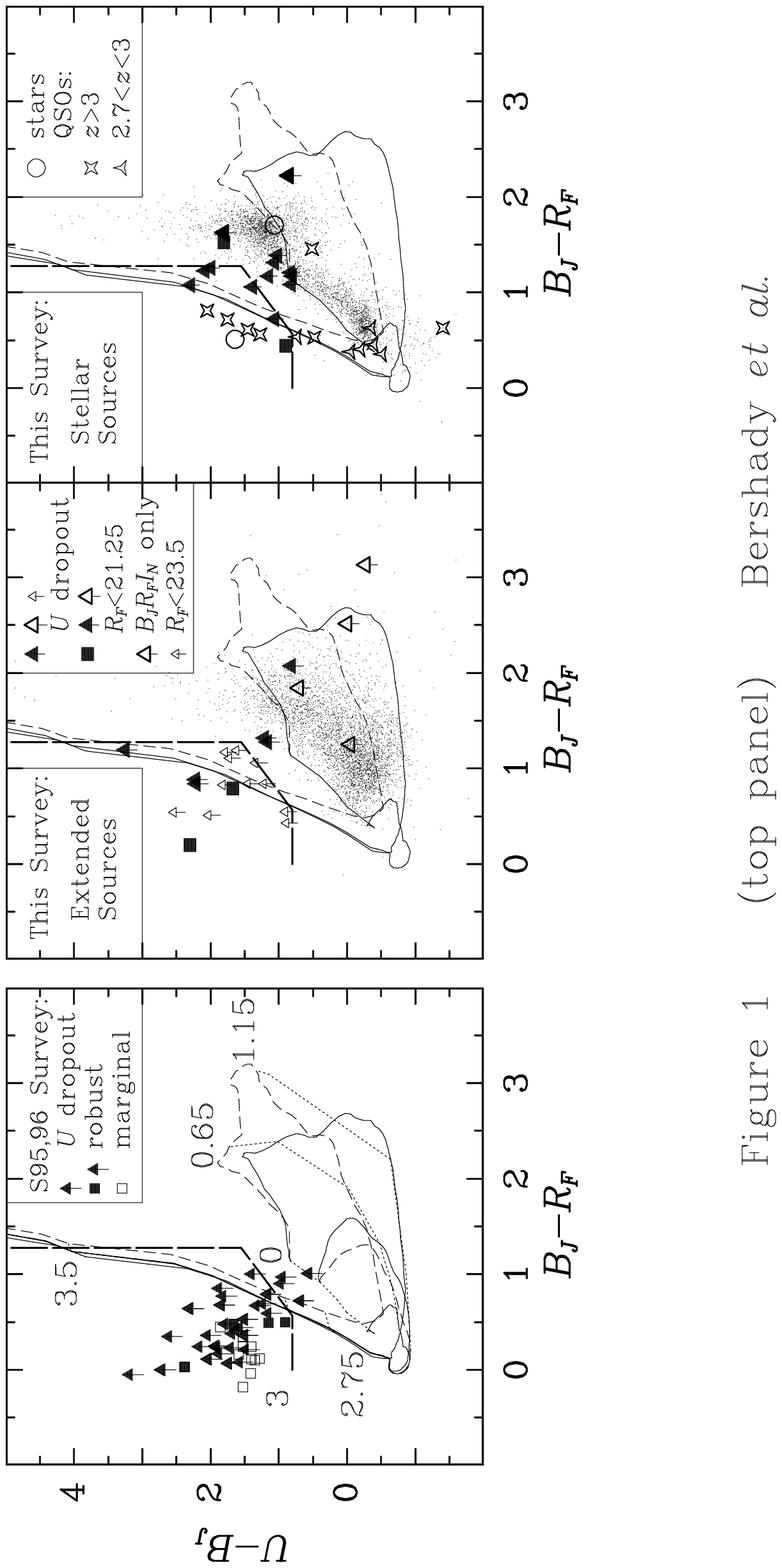}{6in}{0}{90.}{90.0}{-270}{-200}
\vskip 2.15in
\end{figure}

\clearpage

\begin{figure}
\plotfiddle{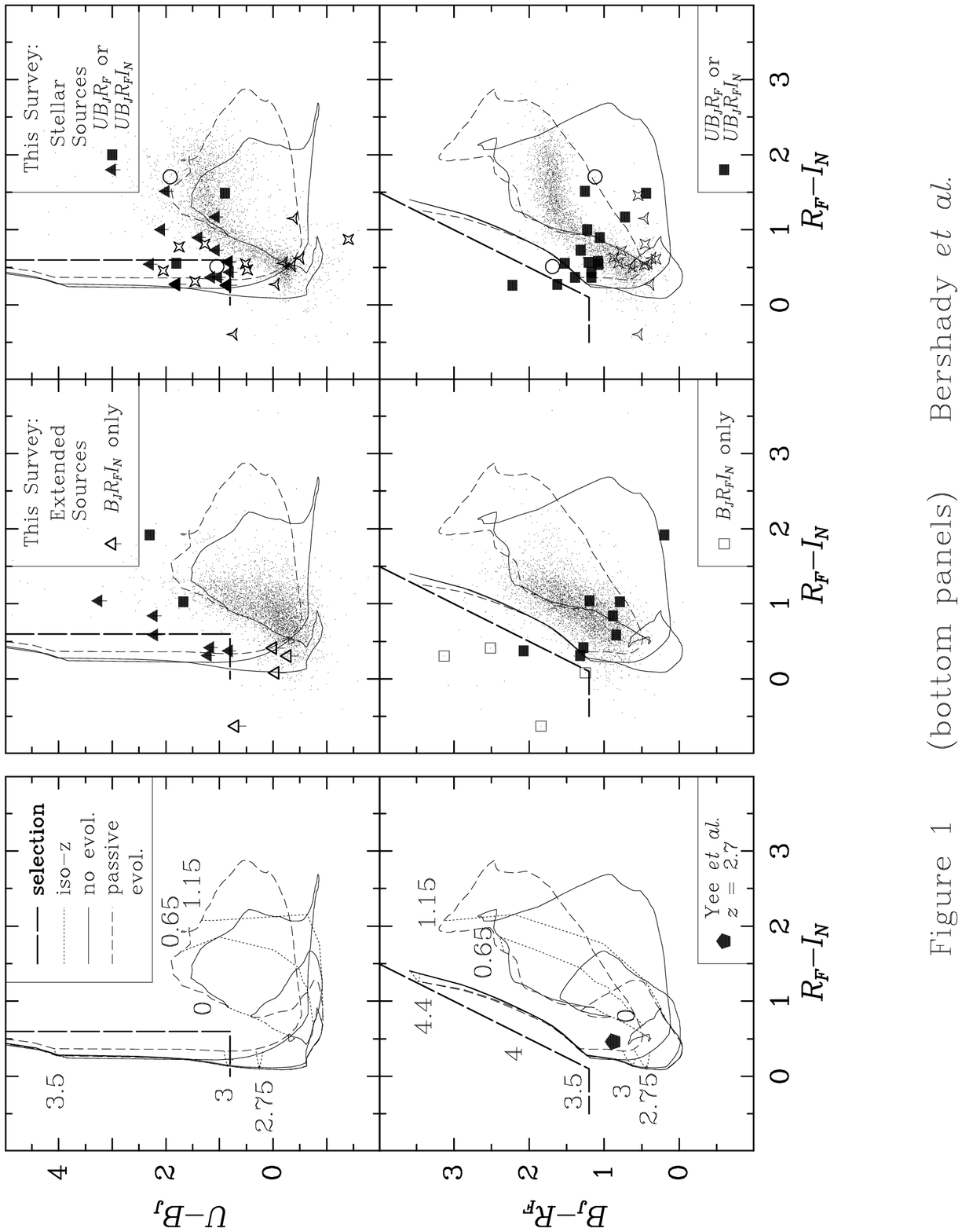}{6in}{0}{90.}{90.0}{-270}{-200}
\vskip 2.15in
\end{figure}

\clearpage

\begin{figure}
\plotfiddle{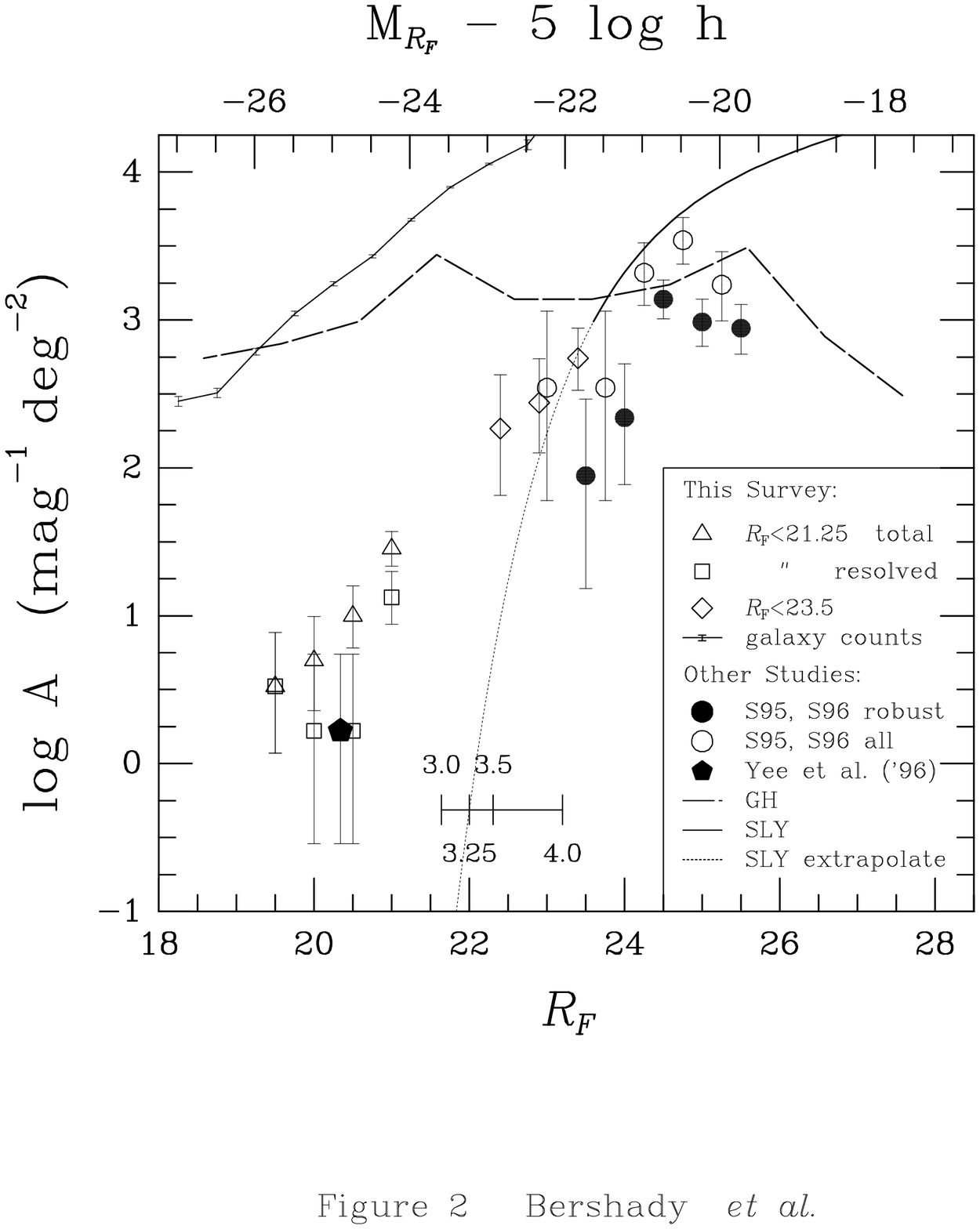}{6in}{0}{100.}{100.0}{-290}{-220}
\vskip 1.0in
\end{figure}


\begin{references}

\reference{} Bruzual, A. G. \& Charlot, S. 1993, ApJ, 405, 538

\reference{} Connolly, A. J., Csabai, I., Szalay, A. S., Koo, D. C.,
Kron, R. G.  1996, AJ, 110, 2655

\reference{} Cowie, L. L. 1988, in ``The Post-Recombination Universe'',
eds. N. Kaiser \& A. Lasenby, (Kluwer: Dordrecht), 1

\reference{} Cowie, L. L., Lilly, S. J., Gardner, J. \& McLean,
I. S. 1988, ApJ, 332, 59

\reference{} Fukugita, M., Shimasaku, K. \& Ichikawa, T., 1995, PASP, 107, 945

\reference{} Gehrels, N. 1986, ApJ, 303, 336

\reference{} Glazebrook, K., Ellis, R. S., Colless, M., Broadhurst, T.,
Allington-Smith, J. \& Tanvir, N. 1995, MNRAS, 273, 157

\reference{} Guhathakurta, P., Tyson, J. A. \& Majewski, S. R. 1990, ApJ, 357, L9

\reference{} Gwyn, S. D. J. \& Hartwick, F. D. A. 1996, ApJ, 468, L77 (GH)

\reference{} Koo, D. C. 1985, AJ, 90, 418

\reference{} Koo, D. C. \etal 1996, ApJ, 469, 535

\reference{} Kron, R. G. 1980, ApJS, 43, 305

\reference{} Kron, R. G., Bershady, M. A., Munn, J. A., Smetanka, J. J.,
Majewski, S. R. 1992, in {\it Workshop on the Space Distribution of
Quasars}, ed. D. Crampton (ASPCS), 21, 32

\reference{} Lowenthal, J. D. \etal 1997, ApJ, 481, 673

\reference{} Madau, P. 1995, ApJ, 441, 18

\reference{} Majewski, S.~R. 1988, in ``Towards Understanding Galaxies at High
Redshift'', eds. R. G. Kron \& A. Renzini, (Kluwer), 203

\reference{} Majewski, S.~R. 1989, in ``The Epoch of Galaxy Formation'',
eds. C.~S. Frenk {\it et al.}, (Kluwer), 85.

\reference{} Majewski, S.~R. 1992, ApJS, 78, 87

\reference{} Majewski, S.~R., Kron, R.~G., Koo, D.~C. \& Bershady,
M.~A. 1994, PASP, 106, 1258

\reference{} Mobasher, B., Rowan-Robinson, M., Georgakakis, A. \& Eaton, N.
1996, MNRAS, 282, L7

\reference{} Munn, J.~A., Koo, D.~C., Kron, R.~G., Bershady, M.~A.,
Majewski, S.~R., Smetanka, J.~J. 1997, ApJS, 109, 45

\reference{} Partridge, B. \& Peebles, P.J. 1967, ApJ, 147, 868

\reference{} Sawicki, M.~J., Lin, H. \& Yee, H.~K.~C. 1997, AJ, 113, 1 (SLY)

\reference{} Steidel, C.~C. \& Hamilton, D. 1993, AJ, 105, 2017

\reference{} Steidel, C. C.,  Pettini, M. \& Hamilton, D. 1995, AJ, 110, 2519 (S95)

\reference{} Steidel, C. C., Giavalisco, M.,  Dickinson, M., \&
Adelberger, K. L. 1996a, AJ, 112, 352.

\reference{} Steidel, C. C., Giavalisco, M., Pettini, M., Dickinson, M., \&
Adelberger, K. L. 1996b, ApJ, 462, L17 (S96)

\reference{} Tyson, J. A. 1988, AJ, 96, 1

\reference{} Yee, H. K. C., Ellingson, E., Bechtold, J., Carlberg, R. G. \&
Cuillandre, J.-C.  1996, AJ, 462, 32

\end{references}
\end{document}